# Observation of plasmon-phonons in a metamaterial superconductor using inelastic neutron scattering

Vera N. Smolyaninova[1], Jeffrey W. Lynn[2], Nicholas P. Butch[2], Heather Chen-Mayer[2], Joseph C. Prestigiacomo[3], M. S. Osofsky[3], and Igor I. Smolyaninov[4,5]

[1]*Department of Physics Astronomy and Geosciences, Towson University,*

*8000 York Rd., Towson, MD 21252, USA*

[2] *NIST Center for Neutron Research, National Institute of Standards and Technology,*

*100 Bureau Drive, Gaithersburg, MD 20899-6102, USA*

[3]*Naval Research Laboratory, Washington, DC 20375, USA*

[4]*Department of Electrical and Computer Engineering, University of Maryland, College Park, MD 20742, USA*

[5] *Saltenna LLC, 1751 Pinnacle Drive #600 McLean, VA 22102, USA*

A metamaterial approach is capable of drastically increasing the critical temperature, $T_c$, of composite metal-dielectric superconductors as demonstrated by the tripling of $T_c$ that was observed in bulk Al-Al$_2$O$_3$ core-shell metamaterials. A theoretical model based on the Maxwell-Garnett approximation provides a microscopic explanation of this effect in terms of electron-electron pairing mediated by a hybrid plasmon-phonon excitation. We report the first observation of this excitation in Al-Al$_2$O$_3$ core-shell metamaterials using inelastic neutron scattering. This result provides support for this novel mechanism of superconductivity in metamaterials.

Recent theoretical [1-3] and experimental [4-6] work has demonstrated that many tools developed in electromagnetic metamaterial research can be used to engineer artificial metamaterial superconductors having improved superconducting properties. This connection between electromagnetic metamaterials and superconductivity research stems from the fact that superconducting properties of a material may be expressed via its effective dielectric response function $\varepsilon_{\text{eff}}(q,\omega)$, and the critical temperature, $T_c$, of a superconductor is defined by the behaviour of $\varepsilon_{eff}^{-1}(q,\omega)$ near its poles [7]. In conventional superconductors these poles are defined by the dispersion law, $\omega(q)$, of phonons which mediate electron-electron pairing. Recently, we have demonstrated a considerable enhancement of the attractive electron-electron interaction in such metamaterial scenarios as epsilon near zero (ENZ) [8] and hyperbolic metamaterials.[9] In both cases the inverse dielectric response function of the metamaterial may exhibit additional poles compared to the parent superconductor. The most striking example of successful metamaterial superconductor engineering was the observation of tripling of the critical temperature $T_c$ in Al-Al$_2$O$_3$ epsilon near zero (ENZ) core-shell metamaterials compared to bulk aluminium [5]. The formation of these bulk Al-Al$_2$O$_3$ samples enable studies that require large sample volumes, such as neutron scattering, that can reveal phonon spectral features that are not accessible in thin films.

Here, we report on the use of inelastic neutron scattering to provide the first experimental evidence of an excitation that does not exist in pure Al and corresponds to the metamaterial pole of the engineered inverse dielectric response function responsible for the $T_c$ enhancement in this material. We also identify the microscopic physical origin of this additional metamaterial pole as coming from a hybrid plasmon-phonon mode which arises in a composite metal-dielectric metamaterial. The hybrid character of this mode enables efficient inelastic neutron scattering from its phonon component, which is observed in the experiment. The direct observation of plasmon-phonon modes in Al-Al$_2$O$_3$ ENZ core-shell metamaterials using inelastic neutron scattering provides strong support of this novel plasmon-phonon mechanism of superconductivity in composite metal-dielectric ENZ metamaterials.

The theoretical model of superconductivity in ENZ metamaterials is based on the paper by Kirzhnits *et al.* [7], which demonstrated that within the framework of macroscopic electrodynamics the electron-electron interaction in a superconductor may be expressed in the form of an effective Coulomb potential

$$V(\vec{q},\omega) = \frac{4\pi e^2}{q^2 \varepsilon_{eff}(\vec{q},\omega)} = \frac{V_C}{\varepsilon_{eff}(\vec{q},\omega)}, \tag{1}$$

where $V_C = 4\pi e^2/q^2$ is the Fourier-transformed Coulomb potential in vacuum, and $\varepsilon_{eff}(q,\omega)$ is the linear dielectric response function of the superconductor treated as an effective medium. The critical temperature of a superconductor in the weak coupling limit is typically calculated as

$$T_c = \theta \; \exp\left(-\frac{1}{\lambda_{eff}}\right), \tag{2}$$

where $\theta$ is the characteristic temperature for a bosonic mode mediating electron pairing (such as the Debye temperature $\theta_D$ in the standard BCS theory) [10]. The dimensionless coupling constant $\lambda_{eff}$ is defined by $V(q,\omega) = V_C \varepsilon^{-1}(q,\omega)$ and the density of states $\nu$ (see for example [11]):

$$\lambda_{eff} = -\frac{2}{\pi}\nu\int\limits_0^\infty \frac{d\omega}{\omega}\left\langle V_C \, \mathrm{Im}\,\varepsilon^{-1}(\vec{q},\omega)\right\rangle, \tag{3}$$

where $V_C$ is the unscreened Coulomb repulsion, and the angle brackets denote average over the Fermi surface. Now this formalism will be applied to a composite metal-dielectric metamaterial.

Following [12], a simplified dielectric response function of a metal may be written as

$$\varepsilon_m\left(q,\omega\right) = \left(1 + \frac{k^2}{q^2}\right)\left(1 - \frac{\Omega^2(q)}{\omega^2 + i\omega\Gamma}\right) \qquad (4)$$

where $k$ is the inverse Thomas-Fermi screening radius, $\Omega(q)$ is the dispersion law of a phonon mode, and $\Gamma$ is the corresponding damping rate. The zero of the dielectric response function of the bulk metal (which occurs at $\omega=\Omega(q)$ where $\varepsilon_m$ changes sign) maximizes the electron-electron pairing interaction given by Eq.(1). This simplified consideration of $\varepsilon_m(q,\omega)$ has been justified in [3].

A superconducting metallic "matrix" with dielectric "inclusions", which forms a composite metal-dielectric metamaterial, is now considered. We will assume that the permittivity $\varepsilon_d$ of the dielectric does not depend on $(q,\omega)$ and stays positive and constant. According to the Maxwell-Garnett approximation [13], mixing of nanoparticles of a superconducting "matrix" with dielectric "inclusions" (described by the dielectric constants $\varepsilon_m$ and $\varepsilon_d$, respectively) results in an effective medium with a dielectric constant $\varepsilon_{eff}$, which may be obtained as

$$\left(\frac{\varepsilon_{eff} - \varepsilon_m}{\varepsilon_{eff} + 2\varepsilon_m}\right) = (1-n)\left(\frac{\varepsilon_d - \varepsilon_m}{\varepsilon_d + 2\varepsilon_m}\right), \qquad (5)$$

where $n$ is the volume fraction of metal ($0 \le n \le 1$). The explicit expression for $\varepsilon_{eff}^{-1}$ may be written as

$$\varepsilon_{eff}^{-1} = \frac{n}{(3-2n)}\frac{1}{\varepsilon_m} + \frac{9(1-n)}{2n(3-2n)}\frac{1}{(\varepsilon_m + (3-2n)\varepsilon_d / 2n)} \qquad . \qquad (6)$$

For a given value of the metal volume fraction $n$, the ENZ conditions ($\varepsilon_{eff} \approx 0$) may be obtained around $\varepsilon_m \approx 0$ (at the phonon frequency $\omega = \Omega(q)$ of the superconducting metal), and around

$$\varepsilon_m \approx -\frac{3-2n}{2n}\varepsilon_d \qquad . \qquad (7)$$

According to Eq.(4), the additional pole of the inverse dielectric response function described by Eq.(7) occurs at some frequency $\omega < \Omega(q)$ close to the phonon resonance. Indeed, Eqs.(4,7) allow us to predict the location of the additional pole with respect to the phonon frequency $\Omega(q)$. Neglecting the unknown value of $\Gamma$ in Eq.(4), we obtain:

$$\varepsilon_m(q,\omega) \approx \left(1+\frac{k^2}{q^2}\right)\left(1-\frac{\Omega^2(q)}{\omega^2}\right) \approx -\frac{3-2n}{2n}\varepsilon_d \qquad . \qquad (8)$$

This leads to the following expression for the ratio of the additional pole and the regular phonon frequencies:

$$\frac{\omega}{\Omega} \approx \left(1+\frac{(3-2n)\varepsilon_d}{2n\left(1+\frac{k^2}{q^2}\right)}\right)^{-1/2} \qquad . \qquad (9)$$

The dielectric constant for $Al_2O_3$ in the long wavelength infrared range is $\varepsilon_d \sim 2.25$. Using the ratio of the inverse Thomas-Fermi radius to the Fermi radius for aluminium $k/k_F = 1.17$ tabulated in [12], and assuming $q = 2k_F$, the $\omega/\Omega$ ratio given by Eq.(9) may be calculated as a function of n. This function is shown in Fig.1. Based on these estimates, we may expect that the additional pole must be located in the $0.4\Omega$-$0.7\Omega$ range depending on the volume fraction of aluminum in the metamaterial.

To search for the existence of this additional excitation, which should occur at frequencies lower than the phonon frequencies of aluminium, an inelastic neutron scattering experiment was performed. To prepare the samples, commercial (US Research Nanomaterials) 18 nm diameter aluminium nanoparticles were oxidized under ambient conditions. Upon oxidation, an approximately 2 nm thick layer of $Al_2O_3$ is formed on the surface of aluminium nanoparticles [14]. The particles were subsequently compressed into dense pellets in a hydraulic press. A SEM image of such a core-shell $Al$-$Al_2O_3$ compressed sample is shown in Fig. 2(a). The superconducting transition of this bulk metamaterial sample was measured to be $T_c = 3.7$ K, which is more than three times higher than $T_c = 1.2$ K of bulk aluminium (Fig. 2 (b)). The zero-field cooled magnetization data shown in Fig. 2(b) were measured using MPMS SQUID magnetometer in magnetic field of 1 mT. Note that the superconducting coherence length in aluminum (which roughly corresponds to the size of a Cooper pair) is known to be very large ($\xi = 1600$ nm – see for example ref. [12]). The superconducting coherence length in the aluminum based hyperbolic metamaterial superconductors has been experimentally measured in [6]. The measured value appears to be $\xi = 181$ nm, which is smaller than $\xi$ in pure aluminium. The coherence length for our $Al$-$Al_2O_3$ ENZ core-shell metamaterial was determined from a measurement of the critical field, $H_{c2}$ (Fig. 3). Although pure Al is a type I superconductor, the granular samples exhibit type II behaviour. Thus, the coherence length, as determined by $\xi = \sqrt{\phi_0 / 2\pi H_{c2}}$ , is 105 nm for the measured $H_{c2}$ of 300 G. It is important to note that this experimentally measured value of the superconducting coherence length is much larger than the 18 nm aluminum grain size, thus confirming the consideration of our samples as "superconducting metamaterials".

The inelastic neutron scattering experiments were performed at the NIST Center for Neutron Research using triple-axis spectrometers BT-7 [15] and BT-4 and the Disk Chopper Spectrometer [16]. For the measurements on BT-7 we employed horizontal focusing with a fixed final energy $E_f = 14.7$ meV to increase the range of the wave vector integration. A pyrolytic graphite filter was used in the scattered beam and a velocity selector in the incident beam to suppress higher-order wavelength contaminations and reduce background [15]. Data were collected in the energy range from 3 to 43 meV at 5K employing a closed cycle helium refrigerator. The high energy data were collected on BT-4 Filter Analyzer Neutron Spectrometer using the Cu(220) monochromator. Data were obtained from 36 meV to 250 meV at 78 K in a liquid nitrogen cryostat. Background data were obtained with the sample lifted out of the beam area. The highest resolution data were collected on the Disk Chopper Spectrometer [16] in a pumped $^4$He cryostat. Data were collected at 1.5 K and 6 K, below and above the $T_c$ of 3.7 K respectively, with an incident energy of 13 meV. In subtracting the 6 K data from the 1.5 K data, we found a small but noticeable increase in the scattering in the superconducting state over a broad energy range from just above the elastic line to $\approx$ 8 meV. This behaviour is typical for inelastic scattering experiments with superconductors.

The results from the inelastic neutron scattering measurements at $T = 5$ K are shown in Fig. 4. The sample was thermally treated at $110^o$C in vacuum before the measurements to remove possible water adsorbed on the particles. Data above 40 meV are dominated by scattering from hydrogen in the form of OH [17], which may come from a small amount of AlOOH. Prompt gamma-ray neutron activation analysis [18,19] showed a hydrogen content of 12 atomic percent H in the sample. The neutron scattering cross section for H is 80.27 compared to Al of 1.5, which is why the H can contribute to the scattering at high energies even though the amount of H is relatively small.

Fig. 4(b) shows averaged data for energies below 43 meV, taken at wave vectors Q=4 Å$^{-1}$, 4.25 Å$^{-1}$, and 4.5 Å$^{-1}$, which are proportional to the generalized phonon density of states (PDOS), which is the phonon density-of-states weighted by the neutron cross sections. The PDOS for bulk aluminum at $T = 10$ K (black circles) [20] is shown for comparison. Two peaks in the aluminium PDOS corresponding to transverse and longitudinal acoustical phonons are also present in core-shell Al-Al$_2$O$_3$ metamaterial, which are indicated with black arrows in Fig. 4(b). However, there is an additional peak at around 15 meV (indicated with blue arrow), which is not present in pure aluminum. This peak cannot be attributed to the aluminum oxide, since aluminum oxide has a stiffer lattice [21], and will have a peak in the PDOS at energies higher than that for aluminium. It also cannot be attributed to the effect of 12% hydroxide impurities. Due to the much smaller mass of hydrogen atoms compared to aluminum, the energy range of inelastic neutron scattering peaks associated with the O-H bonds must be located at much higher energies. The considerable enhancement of inelastic neutron scattering at energies above 40 meV is indeed observed in the experimental results presented in Fig.4(a). Consequently, the inelastic neutron scattering features near 15 meV cannot be affected by hydroxide impurities. Therefore, we have observed an additional contribution to the PDOS which was predicted by our metamaterial model at energies within the span of phonon energies of aluminium. The dispersion of this additional excitation may be evaluated based on the individual inelastic neutron scattering data (Fig. 5) taken on BT-7 at neutron wave vectors Q = 4 Å$^{-1}$, 4.25 Å$^{-1}$, and 4.5 Å$^{-1}$, respectively, as marked in each plot. The additional peak at around 15 meV (indicated with arrows) exhibits very weak dispersion, which is consistent with the hypothesized van-Hove character of this peak.

This peak is also consistent with unexplained extra spectral weight seen at low energy in the $\alpha^2F(\omega)$ extracted from the tunnelling conductance of granular Al films that is absent in pure Al films [22]. Since this quantity is a direct measure of the boson mechanism responsible for superconductivity, it is reasonable to infer that the observed peak reflects a contribution to mechanism responsible for the enhanced $T_c$ in our metamaterial samples. We should also recall that in 1968, Cohen and Abeles observed that the superconductive transition temperature, $T_c$, of thin films of granular aluminium (Al grains coated with $Al_2O_3$) was significantly higher than that of bulk Al (~1.2K) [23]. Several mechanisms were proposed to explain this enhancement but none has proven to be satisfactory. Our observations provide potential explanation for the enhanced $T_c$ in granular aluminium thin films.

Now we discuss the microscopic physical origin of these additional excitations. According to Eq. (7), at $n = 0.75$ the resonant conditions occur at $\varepsilon_m = -\varepsilon_d$, which corresponds to the dispersion law of surface plasmons propagating along planar metal-dielectric interfaces [24]:

$$k = \frac{\omega}{c}\left(\frac{\varepsilon_m \varepsilon_d}{\varepsilon_m + \varepsilon_d}\right)^{1/2} \qquad . \qquad (10)$$

For non-planar interfaces of aluminium nanoparticles the plasmon resonance occurs at slightly different points. However, Eqs. (7, 10) clearly relate the additional pole of the inverse dielectric response function $\varepsilon_{\text{eff}}^{-1}$ of the metamaterial to the plasmon resonance at metal-dielectric interfaces. Since this additional plasmon resonance of the metamaterial structure may exist only in the vicinity of the metal phonon at $\omega = \Omega(q)$ (where $\varepsilon_m$ changes sign), the proper name of this resonant excitation should probably be chosen as a "hybrid" plasmon-phonon resonance. Unlike regular plasmons in metals (which typically occur in the visible and infrared frequency ranges) the hybrid plasmon-phonon mode in metal-dielectric composite metamaterials occurs near the frequency range of conventional phonons. This means that hybrid plasmon-phonons are coupled to lattice vibrations via the lattice polarizability, which enables their detection via inelastic neutron scattering.

Based on Eqs. (2, 3), it is clear that the existence of an additional plasmon-phonon pole of the inverse dielectric response function $\varepsilon_{eff}^{-1}$ of the metamaterial may lead to increased $T_c$. The microscopic physical origin of this effect may be illustrated by Fig. 6(a). As pointed out in [25], plasmon-mediated pairing of electrons may be understood in terms of image charge-mediated Coulomb interaction. Let us consider two electrons located next to a planar interface between two media with dielectric permittivities $\varepsilon_1$ and $\varepsilon_2$, as shown in Fig. 6(a). The field acting on the charge $e_2$ in the medium $\varepsilon_1$ at $z > 0$ is obtained as a superposition of fields produced by the charge $e_1$ and its image $e_1'$ [26]. As a result, the effective Coulomb potential may be obtained as

$$V = \frac{e}{\varepsilon_1}\left( \frac{e}{r_1} - \frac{e}{r_2}\left( \frac{\varepsilon_2 - \varepsilon_1}{\varepsilon_2 + \varepsilon_1} \right) \right), \tag{11}$$

which may be simplified as

$$V = \frac{2e^2}{r(\varepsilon_2 + \varepsilon_1)} \tag{12}$$

if both charges are located very close to the interface, so that $r_1 = r_2 = r$. The $\varepsilon_2 = -\varepsilon_1$ condition, which maximizes the electron pairing interaction, corresponds to the dispersion law of surface plasmons propagating along the interface (see Eq. (10)).

The values of $\lambda_{eff}$ due to each pole in Eq. (6) may be evaluated in terms of $\lambda_m$ for the bulk metal, assuming that $\nu$ in Eq. (3) is proportional to $n$, which is justified by the fact that there are no free charges in the dielectric phase of the metamaterial. We assume that Im($\varepsilon_d$)~0 and neglect the dispersion of Im($\varepsilon_m$) in this simplified estimate (the detailed calculations may be found in [3]). Near the plasmon-phonon pole

$$\text{Im}\left(\varepsilon^{-1}{}_{eff}\right) \approx \frac{9(1-n)}{2n(3-2n)\varepsilon''{}_m} \qquad (13)$$

where $\varepsilon_m'' =$ Im($\varepsilon_m$). Therefore, using Eq.(3), we may obtain the expression for $\lambda_{eff}$ as a function of $\lambda_m$ and $n$ for the plasmon-phonon pole:

$$\lambda_{eff} \approx \frac{9(1-n)}{2(3-2n)}\lambda_m, \qquad (14)$$

where $\lambda_m$ for the parent superconductor is determined by Eq. (3) in the $n \rightarrow 1$ limit. On the other hand, near the regular phonon pole the inverse dielectric response function of the metamaterial behaves as

$$\text{Im}\left(\varepsilon_{eff}^{-1}\right) \approx \frac{n}{(3-2n)\varepsilon_m''} \qquad . \qquad (15)$$

Therefore, near this pole

$$\lambda_{eff} \approx \frac{n^2}{(3-2n)}\lambda_m \qquad . \qquad (16)$$

Assuming the known values $T_{cbulk}$ = 1.2 K and $\theta_D$ = 428 K of bulk aluminium [27], Eq.(2) results in $\lambda_m = 0.17$, which corresponds to the weak coupling limit. Let us plot the hypothetical values of $T_c$ as a function of metal volume fraction $n$, which would originate from either Eq. (14) or Eq. (16) in the absence of each other. The corresponding values calculated as

$$T_c = T_{Cbulk} \exp\left(\frac{1}{\lambda_m} - \frac{1}{\lambda_{eff}}\right) \qquad (17)$$

are shown in Fig. 6b. The vertical dashed line corresponds to the assumed critical value of the metal volume fraction $n_{cr}$ at which the additional plasmon-phonon pole of $\varepsilon_{eff}^{-1}$ disappears as $n \rightarrow 0$ (since according to Eq. (4) the magnitude of $\varepsilon_m$ is limited). The blue dashed line shows the predicted behaviour in the presence of both poles [3]. The experimentally measured data points from [5] are shown for comparison on the same plot. The match between the experimentally measured values of enhanced $T_c$ and the theoretical curve obtained based on Eq. (17) is impressive, given the fact that our model does not contain any free parameters. It implies that the plasmon-phonon mediated electron pairing is the physical mechanism responsible for the tripling of $T_c$ in the Al-$Al_2O_3$ core-shell metamaterial superconductor.

At first glance, the strength of the observed additional peak at ~15 meV in the inelastic neutron scattering data in Fig. 4(b) appears to be quite weak. In fact, the observation of this peak requires quite strong electron-plasmon coupling as discussed below. For the case of conventional electron-phonon superconductors the interaction is very strongly dependent on which part of the Brillouin zone is being studied. Typically such anomalies are only observed when measuring single crystals, even in systems that exhibit strong phonon anomalies (which is not the case for pure Al). Furthermore, one has to know where to look. A good example is $YNi_2B_2C$ [28], which has an enormous anomaly at one particular wave vector. In that case, the Fermi wave vector is revealed by the incommensurate magnetic ordering in the related $ErNi_2B_2C$ and $HoNi_2B_2C$ magnetically ordered superconductors [29], so one knows where to look for a large electron-phonon interaction. On the other hand, when measuring the (generalized) phonon density of states (GPDOS) such as for the present nanoparticle system, any phonon anomaly gets averaged with all the rest of the phonons and most often only a very small anomaly, or no anomaly, is seen. That is certainly the case for pure aluminum. Therefore, the only way one would see any effect in the described system is if there is a strong coupling when the plasmon and phonon dispersions cross, since for inelastic neutron scattering one only sees the effects through the change in the phonons (since neutrons do not "see" the plasmons or electrons directly).

It would also be inappropriate to compare the size of the observed anomaly in the GPDOS with the van Hove singularity for the transverse acoustic phonons in the phonon density of states around 20 meV, where these phonons become dispersionless at the zone boundary. Typically, it is difficult to see any effect in a phonon density of states measurement since the effect only occurs in a small region of (Q,E), no matter how strong the coupling is, as the GPDOS averages everything. The fact that we do see an anomaly is interesting and requires quite strong coupling.

This work was supported in part by the DARPA Award No:W911NF-17-1-0348 "Metamaterial Superconductors", by ONR through the NRL Basic Research Program, and NSF MRI Award No:1626326. The identification of any commercial product or trade name does not imply endorsement or recommendation by the National Institute of Standards and Technology.

**Figure Captions**

**Figure 1**. Ratio of the plasmon-phonon and the phonon frequencies calculated as a function of metal volume fraction $n$ in the Al-Al$_2$O$_3$ metamaterial using Eq. (9). The arrow indicates the frequency of an additional peak observed in the neutron scattering experiment.

**Figure 2.** (a) Scanning Electron Microscope image of the Al-Al$_2$O$_3$ metamaterial. (b) Temperature dependence of zero field cooled magnetization per unit mass for fresh and oxidized Al-Al$_2$O$_3$ core-shell metamaterial samples made of 18 nm aluminium nanoparticles. The observed onset of superconductivity at $\sim$ 3.7 K in the oxidized sample is 3.25 times larger than $T_c$ = 1.2 K of bulk aluminium [5]. The inset shows photo of the compressed metamaterial sample (note: 1 emu $= 10^{-3}$ A m$^2$).

**Figure 3**. Magnetization as a function of magnetic field for the core-shell Al-Al$_2$O$_3$ metamaterial at $T$ = 1.75 K. The paramagnetic background is subtracted from the data.

**Figure 4.** (a) Inelastic neutron scattering data obtained for the core-shell Al-Al$_2$O$_3$ metamaterial at energies below 250 meV. The lower-energy data were obtained on BT-7 and the high energy data were taken on BT-4 FANS. Data above 40 meV are dominated by scattering from hydrogen in the form of OH, which may come from a small amount of AlOOH in the metamaterial. (b) Inelastic neutron scattering data for energies below 43 meV taken on BT-7, averaged over measurements taken at wave vectors Q = 4 Å$^{-1}$, 4.25 Å$^{-1}$, and 4.5 Å$^{-1}$. Uncertainties originate from counting statistics and correspond to one standard deviation. The data were taken at T = 5 K. The averaged dependence is proportional to the generalized phonon density of states (PDOS). The DOS for bulk aluminum at T = 10 K (black circles) [20] is shown for comparison. Two peaks which are indicated with black arrows in the aluminium PDOS correspond to the van-Hove peaks for the transverse and longitudinal acoustical phonons, respectively, at the Brillouin zone boundaries. They are also present in the core-shell Al-Al$_2$O$_3$ metamaterial. The additional peak at around 15 meV (indicated with blue arrow) is not present in pure aluminum. It corresponds to the hybrid plasmon-phonons of the metamaterial.

**Figure 5**. Inelastic neutron scattering data for energies below 43 meV taken on BT-7 for separate neutron wave vectors Q = 4 Å$^{-1}$, 4.25 Å$^{-1}$, and 4.5 Å$^{-1}$, respectively, as marked in each plot. The additional peak at around 15 meV (indicated with arrows) is not present in pure aluminum.

**Figure 6**. (a) Schematic view of the surface plasmon resonance geometry: an electron $e_2$ located next to the interface between two media with dielectric permittivities $\varepsilon_1$ and $\varepsilon_2$, interacts with electron $e_1$ and its image $e_1$'. Resonant conditions are obtained at $\varepsilon_1 = -\varepsilon_2$. (b) Plots of the theoretically calculated values of $T_c$ as a function of metal volume fraction $n$, which would originate from either phonon (Eq. (16), black line) or plasmon-phonon (Eq. (14), red line) pole of the inverse dielectric response function of the Al-Al$_2$O$_3$ core-shell metamaterial in the absence of each other. Blue dashed line shows the predicted behaviour in the presence of both poles. The experimentally measured data points from [5] are shown for comparison on the same plot. The vertical dashed line corresponds to the assumed value of $n_{cr}$.

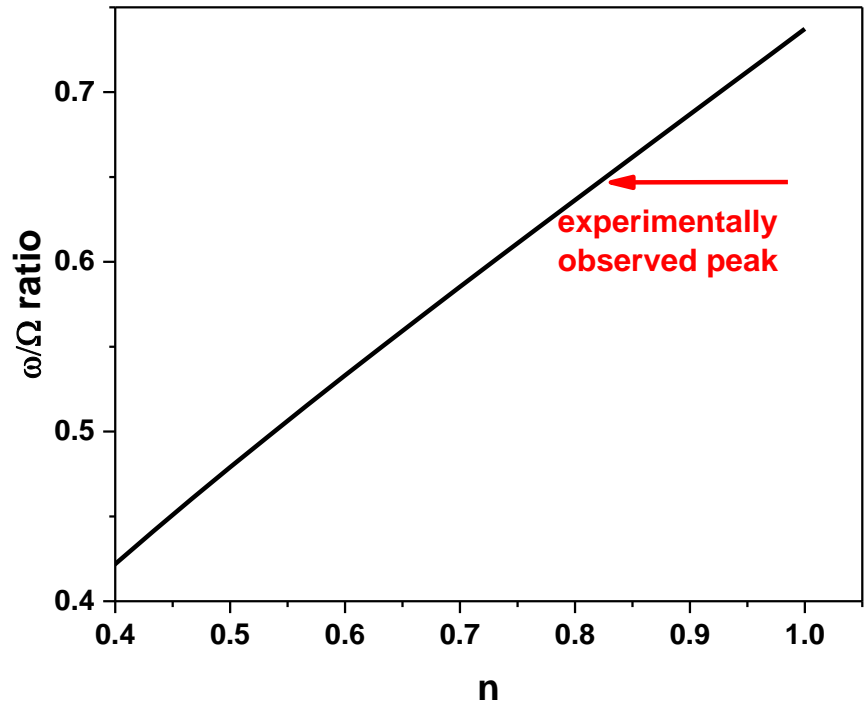

Fig. 1

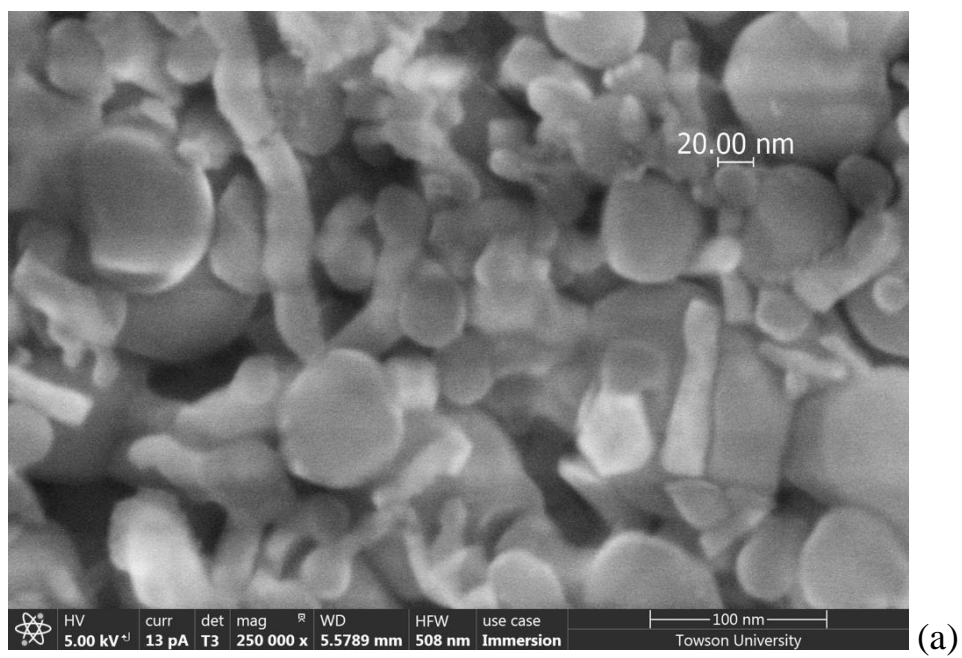

(a)

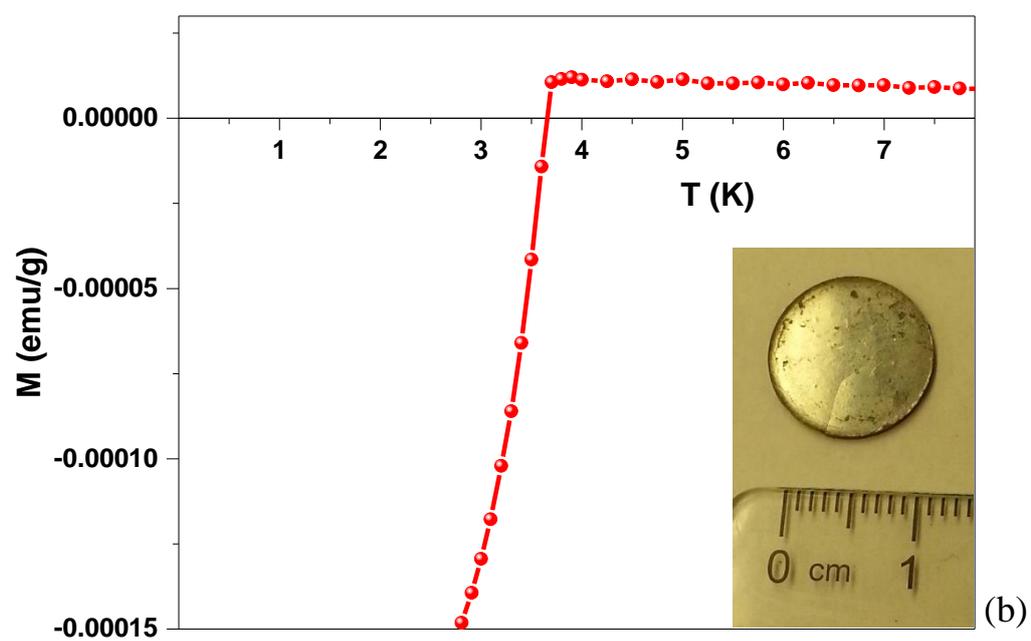

(b)

Fig. 2

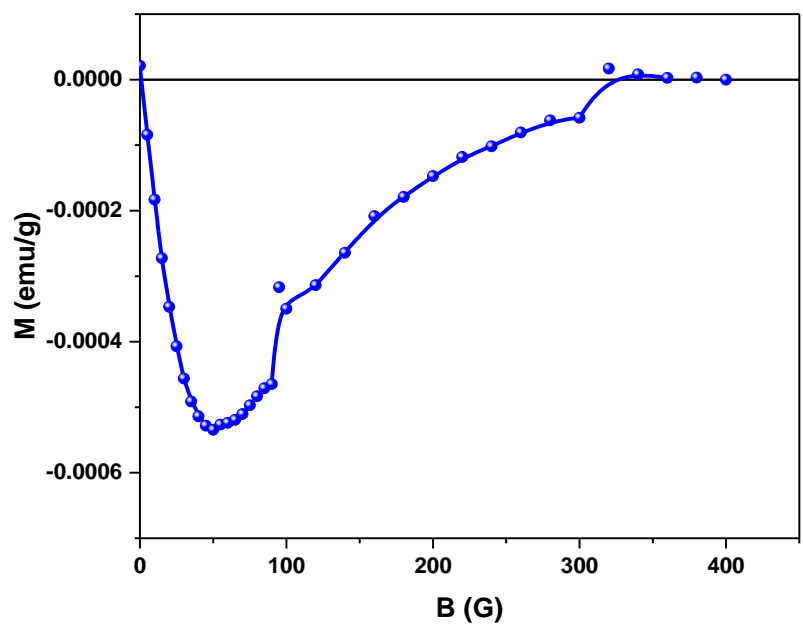

Fig. 3

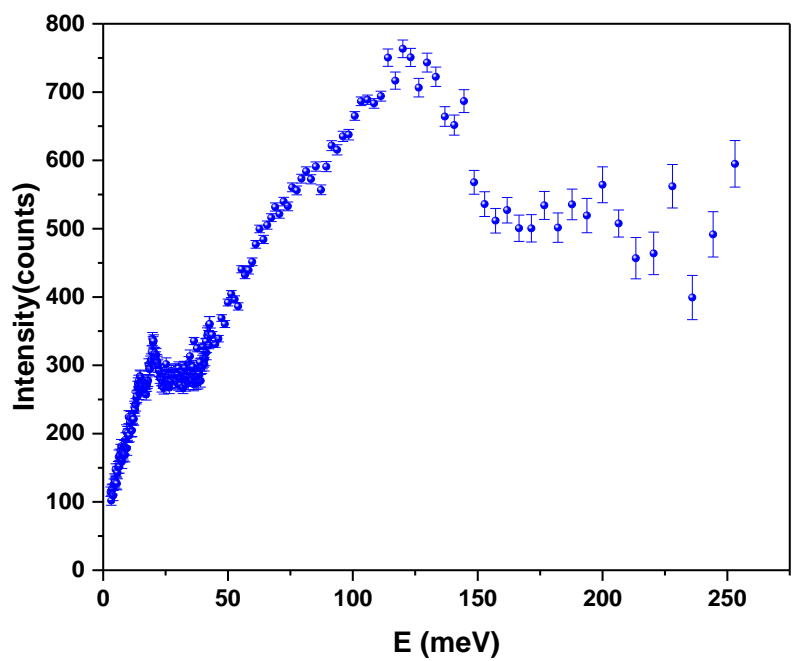

(a)

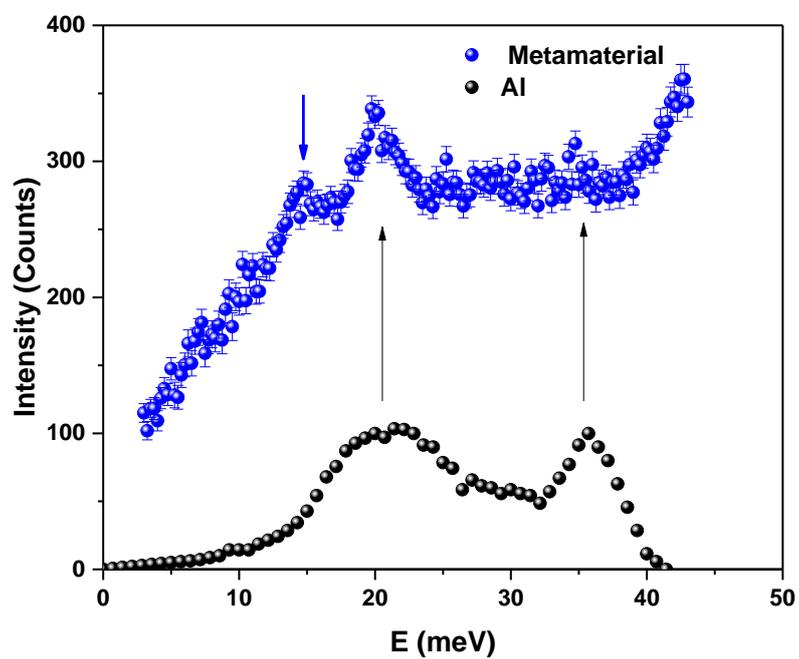

(b)

Fig. 4.

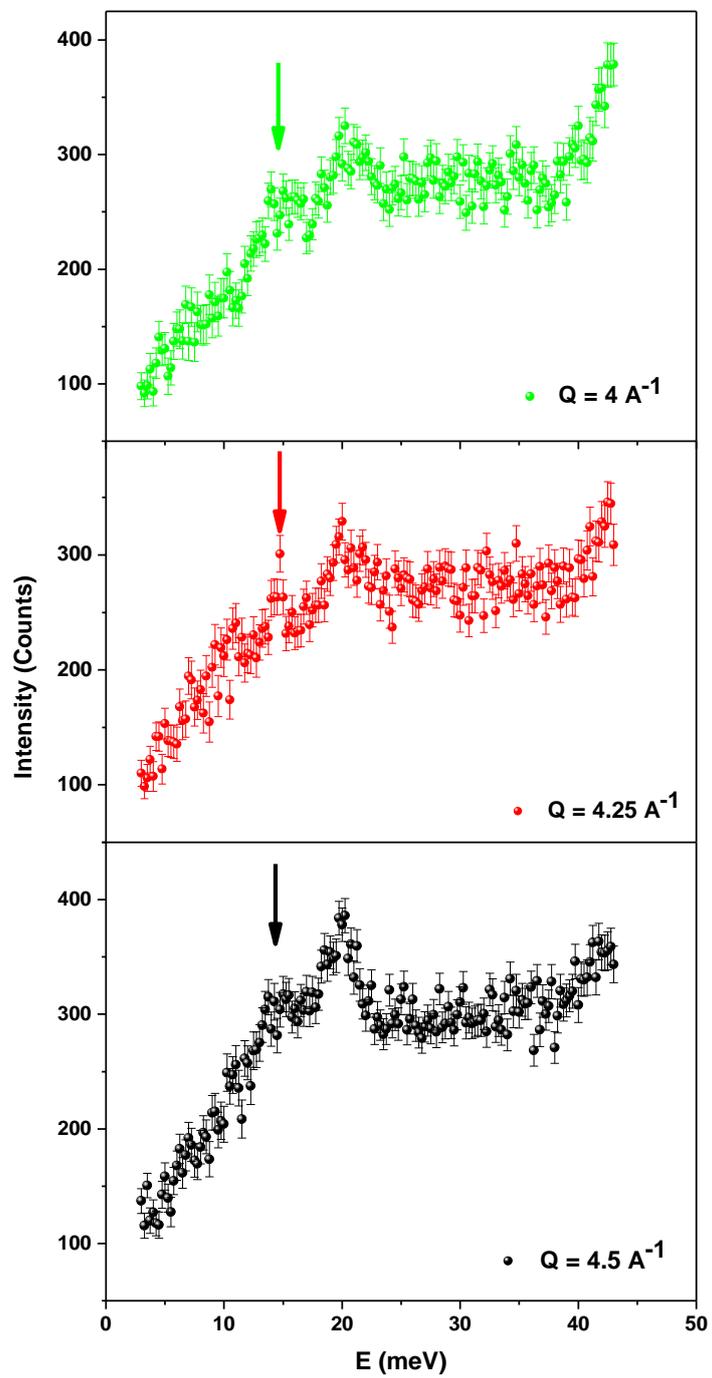

Fig. 5

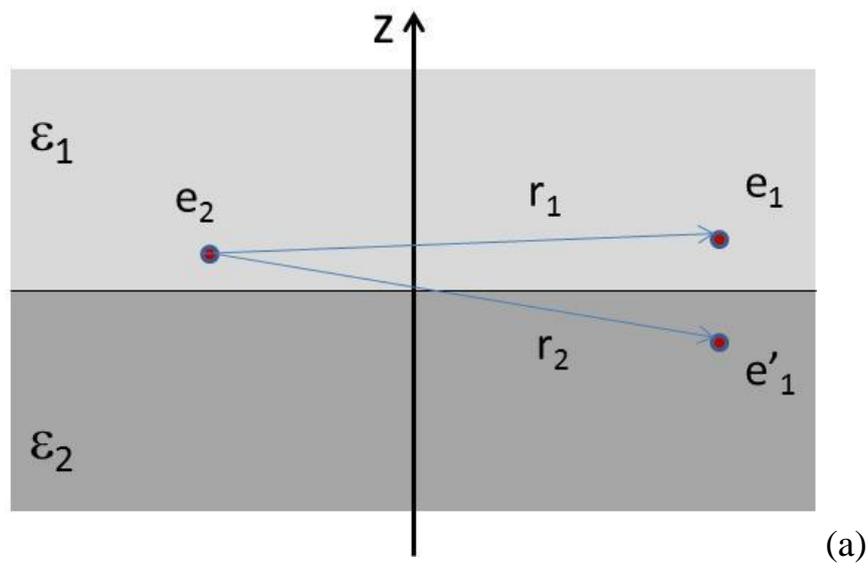

(a)

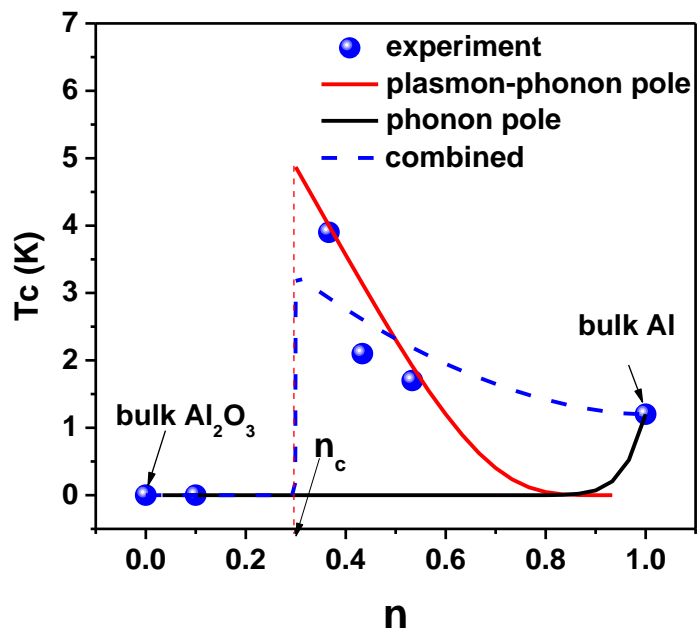

(b)

Fig. 6